\documentclass[prc,twocolumn,floatfix,groupedaddress,nofootinbib,showpacs,preprintnumbers,
amsmath,amssymb,amsfonts,superscriptaddress] {revtex4}
\usepackage{graphicx}
\usepackage{dcolumn}
\usepackage{mathrsfs}
\usepackage{bm}

\usepackage{graphicx}
\usepackage{dcolumn}
\usepackage{bm}
\usepackage[usenames]{color}
\begin{document}

\title{On the compactness of neutron stars}
\author{Wei-Chia Chen}
\email{wc09c@my.fsu.edu} 
\affiliation{Department of Physics, Florida State University, Tallahassee, FL 32306} 
\author{J. Piekarewicz}
\email{jpiekarewicz@fsu.edu}
\affiliation{Department of Physics, Florida State University, Tallahassee, FL 32306}
\date{\today}
\begin{abstract}
Recent progress in the determination of both masses and radii of neutron stars
are starting to place stringent constraints on the dense matter equation of state. 
In particular, new theoretical developments together with improved statistical tools 
seem to favor stellar radii that are significantly smaller than those predicted by 
models using purely nucleonic equations of state. Given that the underlying equation 
of state must also account for the observation of $2M_{\odot}$ neutron stars, 
theoretical approaches to the study of the dense matter equation of state are facing 
serious challenges. In response to this challenge, we compute in a model-independent 
way the underlying equation of state associated with an assumed mass-radius template 
similar to the ``common radius'' assumption used in recent studies. Once such a 
mass-radius template is adopted, the equation of state follows directly from the 
implementation of Lindblom's algorithm; assumptions on the nature or composition 
of the dense stellar core are not required. By analyzing mass-radius profiles with a
maximum mass consistent with observation and common radii in the 8 to 11\,km
range, a lower limit on the stellar radius of a $1.4 M_{\odot}$ neutron star
of $R_{\rm NS}\!\gtrsim\!10.7$\,km is obtained in order to prevent the equation of state 
from violating causality.
\end{abstract}
\maketitle

\emph{How does subatomic matter organize itself and what phenomena 
emerge} is one of the overarching questions guiding the field of nuclear 
physics\,\cite{national2012Nuclear}. In the case of atomic nuclei, the
quest to answer this question requires understanding the nature of the 
nuclear force and the limits of nuclear existence. In the case of extended 
nucleonic matter, this involves elucidating the nature of neutron stars and 
dense nuclear matter. In this letter we focus on the latter.

Owing to the long-range nature of the Coulomb force, extended nucleonic 
matter must be electrically neutral. As a result, dense nuclear matter 
must be by necessity neutron-rich. This is because the electronic 
contribution to the energy increases rapidly with density, so electron 
capture becomes energetically advantageous. Given that such 
extreme conditions of density and isospin asymmetry can not be 
realized in terrestrial experiments, neutron stars have become unique 
laboratories for the exploration of dense matter. This situation has created 
a strong synergy between nuclear physics and astrophysics, that has
been cemented even further through an intimate interplay between theory,
experiment, and observation\,\cite{Horowitz:2014bja}. Indeed, powerful
telescopes operating at a variety of wavelengths drive new theoretical 
and experimental efforts which in turn suggest new observations.

A recent example of such a unique synergy is how accurate 
measurements of massive neutron stars\,\cite{Demorest:2010bx,Antoniadis:2013pzd} 
have informed nuclear models that fall under the general rubric of density functional 
theory. Density functional theory (DFT) offers a comprehensive---and likely 
unique---framework to describe strongly interacting nuclear many-body systems 
ranging from finite nuclei to neutron stars. Rooted on the seminal work by Kohn 
and collaborators\,\cite{Kohn:1999}, DFT shifts the focus from the complicated 
many-body wave function to the much simpler one-body density. The implementation
of DFT to nuclear physics requires that the parameters of the model---which encode 
some of the complicated many-body dynamics---be determined by fitting directly to 
experimental data. In this regard, the accurate measurement of neutron star masses 
has been vital to the accurate calibration of some modern energy density 
functionals\,\cite{Erler:2012qd,Chen:2014sca,Chen:2014mza}.

However, whereas the determination of neutron star masses is accurate and
beyond question, attempts to reliably extract stellar radii\,\cite{Ozel:2010fw,
Steiner:2010fz,Suleimanov:2010th} have been hindered by large systematic 
uncertainties that resulted in an enormous disparity in stellar radii---ranging 
from as low as 8\,km\,\cite{Ozel:2010fw} all the way to 14\,km\,\cite{Suleimanov:2010th}. 
It appears, however, that since those first analyzes were performed, the situation 
has significantly improved in the last few years through a better understanding of 
systematic uncertainties, important theoretical developments, and the implementation 
of robust statistical methods\,\cite{Guillot:2013wu,Lattimer:2013hma,
Heinke:2014xaa,Guillot:2014lla,Ozel:2015fia}. Although a consensus has yet 
to be reached, these recent studies seem to favor stellar radii in the
9--11\, km range. Particularly intriguing among these are the results
by Guillot and collaborators that suggest a ``common radius'' of 
$R_{\rm NS}\!=\!9.1^{+1.3}_{-1.5}\,{\rm km}$ for all five quiescent low 
mass x-ray binaries used in their analysis\,\cite{Guillot:2013wu}; this 
common-radius value has been slightly revised to 
$R_{\rm NS}\!=\!(9.4\pm1.2)\,{\rm km}$\,\cite{Guillot:2014lla}. 
What makes this result especially provocative is that satisfying the small 
radius and large mass constraints simultaneously is enormously challenging. 
Indeed, to our knowledge no optimized energy 
density functional can simultaneously reproduce both of these constraints. 
And from the very large number of models available in the 
literature\,\cite{Lattimer:2006xb}, we are aware of only a few that 
account for both\,\cite{Wiringa:1988tp,Hebeler:2010jx,Hebeler:2013nza}. 
From these, the model due to Wiringa, Fiks, and 
Fabrocini\,\cite{Wiringa:1988tp} relies on a microscopic approach based 
on the Argonne $v_{14}$ (AV14) nucleon-nucleon potential supplemented 
with the Urbana VII (UVII) three-nucleon potential. This model predicts 
a maximum stellar mass of $2.13\,M_{\odot}$ and a ``common radius'' 
of $R_{\rm NS}\!\simeq\!10.4$\,km. The other theoretical approach 
due to  Hebeler and collaborators\,\cite{Hebeler:2010jx,Hebeler:2013nza} 
is also microscopic in nature, but instead uses nuclear interactions derived 
from chiral effective field theory. In particular, their softest equation of state
is consistent with the $\sim\!2M_{\odot}$ limit and predicts a radius
as low as $R_{1.4}\!=\!9.7$\,km for a $1.4M_{\odot}$ neutron 
star\,\cite{Hebeler:2013nza}. 

In this letter we aim to elucidate the nature of neutron star matter by relying 
on a powerful result first proven by Lindblom more than two decades
ago\,\cite{Lindblom:1992}. It is a well known fact that all stellar profiles may 
be determined from the Tolman-Oppenheimer-Volkoff (TOV) equations once 
an equation of state $P\!=\!P({\mathcal E)}$ relating the pressure $P$ to the 
energy density ${\mathcal E}$ is supplied. That is, given an equation of state 
(EOS), the TOV equations 
\begin{subequations}
 \begin{align}
  & \frac{dP(r)}{dr} = -G \frac{\Big({\mathcal E}(r)+P(r)\Big)
      \Big(M(r)+4\pi r^{3}P(r)\Big)}{r^{2}\Big(1-2GM(r)/r\Big)}, \\
  & \frac{dM(r)}{dr} = 4\pi r^{2} {\mathcal E}(r),
  \end{align}
 \label{TOV}
\end{subequations}
can be solved once a value for the central pressure $P(0)\!=\!P_{c}$ and enclosed 
mass $M(0)\!=\!0$ are specified at the origin. In particular, the stellar radius $R$ 
is determined from the condition $P(R)\!=\!0$ and the corresponding stellar mass 
as $M\!=\!M(R)$. In this manner, the EOS generates the mass-radius (MR) 
relationship for neutron stars. What Lindblom was able to prove and implement is 
that the inverse also holds true: knowledge of the MR relation uniquely determines 
the neutron star matter equation of state\,\cite{Lindblom:1992}.   

In the spirit of Lindblom's approach, we now proceed to outline the methodology
required to obtain the unique equation of state associated with an assumed MR 
template that closely resembles the ``common-radius'' hypothesis adopted in 
Ref.\,\cite{Guillot:2013wu}. This choice of template is the only model assumption 
made in this work; no additional assumptions on the nature or composition of 
dense matter are required. For alternative attempts at obtaining generic 
model-independent constraints see Ref.\,\cite{Alford:2015dpa} and references contained 
therein.

\begin{figure}[ht]
\vspace{-0.05in}
\includegraphics[width=0.45\textwidth,angle=0]{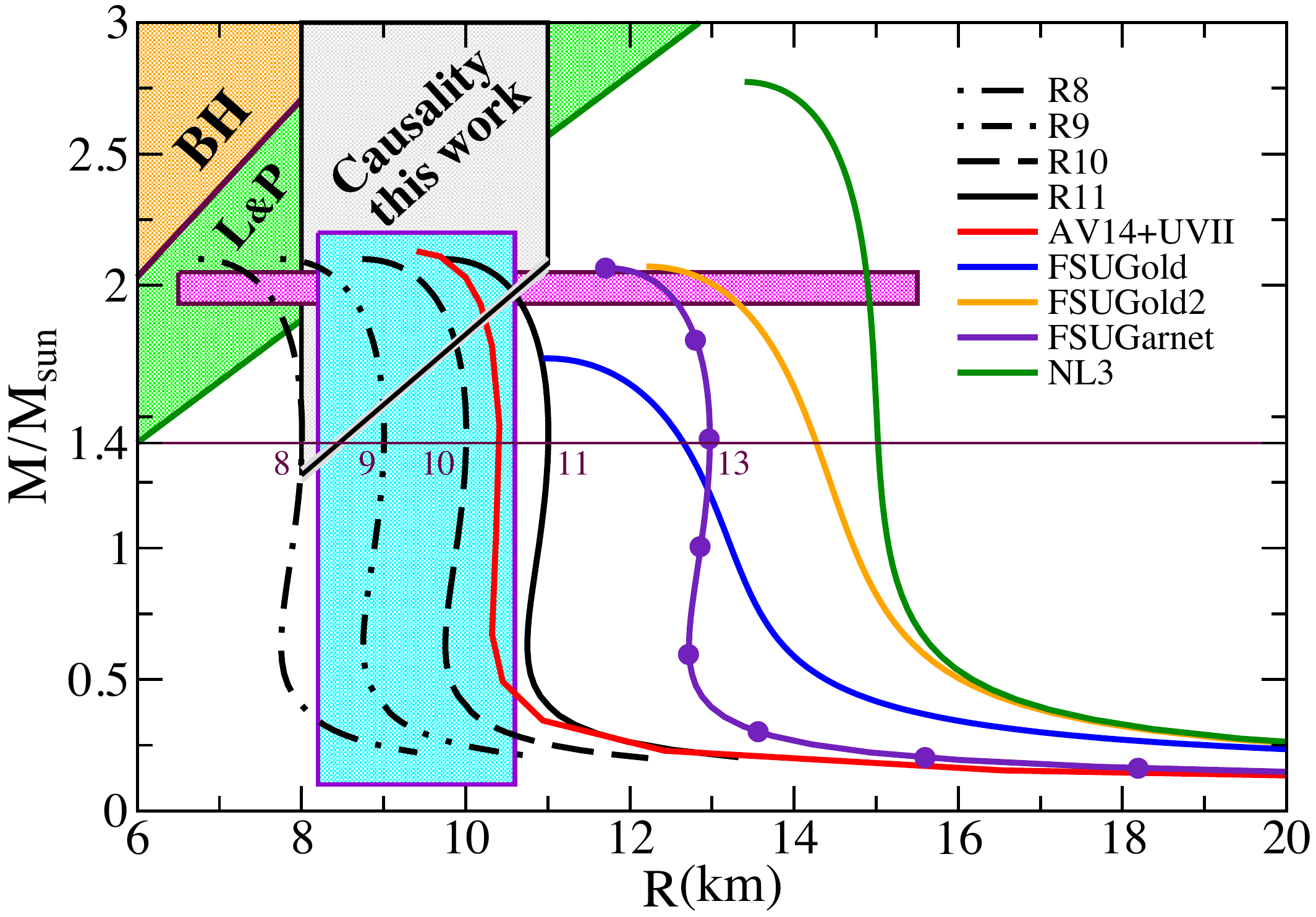}
\caption{(Color online) Mass-Radius profiles as predicted by the
microscopic model of Ref.\,\cite{Wiringa:1988tp} and by the four
relativistic density functionals considered in the text. As a check,
the circles on top of the FSUGarnet profile were obtained using
an EOS extracted from Lindblom's algorithm. Also shown
are the four profiles R8--R11 generated from the FSUGarnet
template and having radii of 8--11\, km, respectively. The horizontal 
band shows the observational constraints from the mass measurements 
reported in Refs.\,\cite{Demorest:2010bx,Antoniadis:2013pzd}, whereas 
the vertical band displays the constraint on stellar radii published
in \,\cite{Guillot:2014lla}. Also shown are regions excluded on purely
theoretical grounds: BH is the black hole limit, L\&P is the
causality constraint obtained in Ref.\,\cite{Lattimer:2006xb}, and 
the grey region defines the causality constraint obtained in this 
work.}
\label{Fig1}
\end{figure}

The particular MR template adopted 
in this work follows from our latest optimization: a relativistic energy density functional 
labeled ``FSUGarnet"\,\cite{Chen:2014mza}. Such a template, among many others  
depicted in Fig.\ref{Fig1}, is displayed with the purple solid line containing a few 
additional circles. FSUGarnet was optimized with inputs from the ground-state 
properties of finite nuclei, their monopole response, and a maximum neutron star 
mass consistent with observation. Yet, even after such an optimization, 
extrapolations to the high-density domain remain highly uncertain. However, it is 
worth underscoring that relativistic density functionals provide a Lorentz 
covariant---and hence causal---framework that becomes critical as one extrapolates 
to the high densities encountered in the stellar interior. Of particular relevance to this 
work are predictions for the maximum neutron star mass and radius of a 1.4$M_{\odot}$ 
neutron star; FSUGarnet predicts: $M_{\rm max}\!=\!(2.07\pm0.02)\,M_{\odot}$ and 
$R_{1.4}\!=\!(13.0\pm0.1)$\,km. Although such a stellar radius appears to significantly 
exceed some of the preferred limits, the predicted EOS is relatively soft in the 
immediate vicinity of nuclear matter saturation density. This is relevant as the 
pressure in the neighborhood of twice nuclear matter saturation density sets 
the overall scale for stellar radii\,\cite{Lattimer:2006xb}. 

Having defined the underlying template, additional MR curves may be easily
generated by simply shifting the full FSUGarnet profile to the desired radius. 
Indeed, the curves labeled R8--R11 in Fig.\,\ref{Fig1} were generated in precisely 
this manner and are characterized by stellar radii of 8--11\,km (rather than 13\,km) 
for a $1.4M_{\odot}$ neutron star. In the case of these four curves, the 
underlying EOS is unknown. Besides the FSUGarnet template, MR predictions
generated with other relativistic density functionals are also included for comparison.
These range from NL3\,\cite{Lalazissis:1996rd,Lalazissis:1999} and 
FSUGold\,\cite{Todd-Rutel:2005fa} that were optimized without incorporating
neutron star information, to the more recent FSUGold2 parametrization that 
includes the $2M_{\odot}$ constraint in the calibration\,\cite{Chen:2014sca}.
Note that all these functionals provide an accurate descriptions of a variety of
ground-state properties of finite nuclei. Finally, displayed with the red line is the
MR relation predicted by the microscopic formalism of  Wiringa, Fiks, and 
Fabrocini\,\cite{Wiringa:1988tp}. With the exception of FSUGold, all MR 
relations are consistent with the $2M_{\odot}$ limit depicted in the figure with 
the narrow magenta band. However, none of the relativistic density functionals 
satisfy the common-radius constraint of Ref.\,\cite{Guillot:2014lla} depicted 
in Fig.\,\ref{Fig1} by the vertical (cyan) band. 

Our next step is to use Lindblom's approach to obtain the EOS corresponding to
the R8--R11 profiles displayed in Fig.\,\ref{Fig1}. Briefly, the ``inversion" procedure 
is implemented as follows. First, one assumes complete knowledge of the EOS up
to a pressure $P_i$ and energy density $\mathcal{E}_{i}$ capable of supporting 
neutron stars up to a mass of about $0.4\,M_{\odot}$. Specifically, using such an 
EOS and a central pressure of $P_{c}\!=\!P_{i}$ generates, through the TOV
equations, a neutron star of mass $M_{i}\!\approx\!0.4\,M_{\odot}$ and radius $R_{i}$. 
Second, one steps along the mass-radius trajectory by selecting a neutron star 
of mass $M_{i+1}\!>\!M_{i}$ and radius $R_{i+1}$ whose central pressure $P_{c}$ 
and energy density $\mathcal{E}_{c}$ are to be determined. To compute $P_{c}$ 
and $\mathcal{E}_{c}$, the TOV equations are integrated inwards from the surface
to the core, with boundary conditions given by $P(R_{i+1})\!=\!0$ and 
$M(R_{i+1})\!=\!M_{i+1}$. Hydrostatic equilibrium guarantees that as one moves
towards the interior of the star, the pressure will continue to increase until the 
maximum known pressure $P_{i}$ will be reached at a certain radius $r_i$ 
located near the center of the star. Given that $r_{i}$ is close to the origin, one
may use suitable series expansions to determine the central pressure $P_{c}$
and energy density $\mathcal{E}_{c}$ in terms of known quantities, i.e., mass, 
pressure, and energy density, at the small radius $r_{i}$\,\cite{Lindblom:1992}. 
In this manner, after one iteration the EOS is extended from $(P_{i}, \mathcal{E}_i)$ 
to $(P_{i+1},\mathcal{E}_{i+1})\!\equiv\!(P_{c}, \mathcal{E}_{c})$. In the next iteration, 
one proceeds in exactly the same fashion, namely, one moves another step along 
the mass-radius trajectory and repeats the algorithm with the newly augmented 
equation of state. After the whole mass-radius trajectory is sampled, the entire 
high-density component of the equation of state is mapped out. Testing the 
reliability and accuracy of Lindblom's inversion algorithm is fortunately very
simple. To start, one uses a known EOS---such as the one predicted by
FSUGarnet---to compute its associated MR relation. Then, one applies 
Lindblom's algorithm to extract the ``new" EOS directly from the MR 
profile. Finally, one verifies that both the new EOS and the resulting 
MR profile are consistent with the originals. The result of such a test
is illustrated in Fig.\ref{Fig1} for the case of FSUGarnet; the circles represent
the results obtained with an EOS extracted from Lindblom's algorithm.

\begin{figure}[ht]
\vspace{-0.05in}
\includegraphics[width=0.45\textwidth,angle=0]{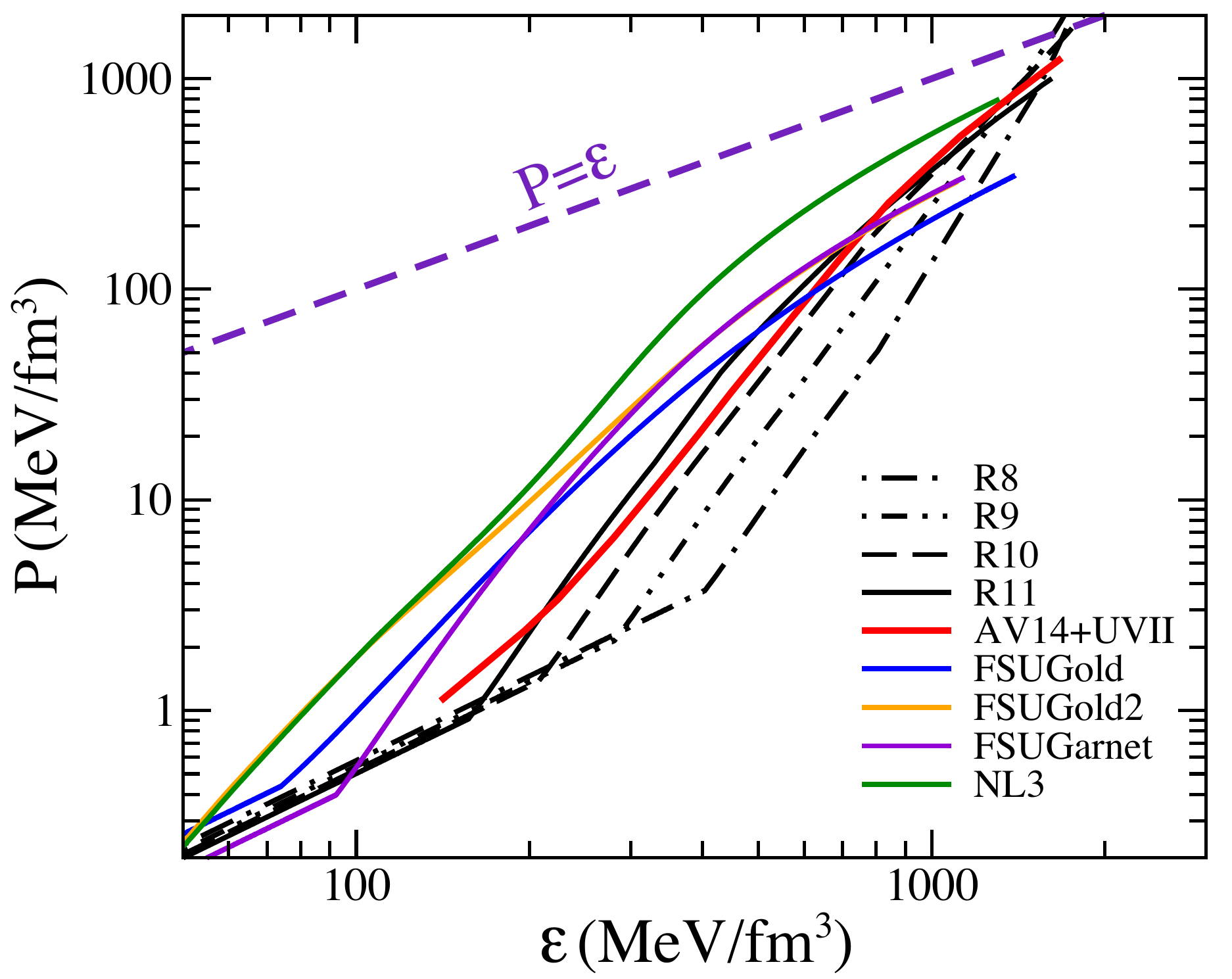}
\caption{(Color online) Equations of state associated to the 
mass-radius profiles displayed in Fig.\,\ref{Fig1}. In the case
of the R8--R11 profiles, the equations of state were obtained
from Lindblom's algorithm\,\cite{Lindblom:1992}. Also shown
is the $P\!=\!\mathcal{E}$ line which demarcates the boundary 
beyond which an equation of state becomes ultrabaric. Equations 
of state that cross the line become superluminal at considerably
lower energy densities.} 
\label{Fig2}
\end{figure}

Having verified the accuracy of Lindblom's algorithm, we can now obtain the 
equations of state that generate the R8--R11 MR profiles assumed in 
Fig.\,\ref{Fig1}. The resultant equations of state along with those predicted
by the non-relativistic and relativistic models are displayed in Fig.\,\ref{Fig2}. 
Note that among the relativistic density functionals, FSUGarnet displays the 
softest EOS at low energy densities
($\mathcal{E}\!\lesssim\!200\,{\rm MeV/fm^{3}}$). For reference, FSUGarnet 
predicts a pressure for pure neutron matter at saturation density of 
$P_{\rm PNM}\!=\!(2.60\pm0.08)\,{\rm MeV/fm^{3}}$. Given that the 
pressure in the vicinity of twice nuclear matter saturation density sets the 
overall scale for stellar radii\,\cite{Lattimer:2006xb}, FSUGarnet generates
the mass-radius profile with the smallest radii; in contrast, NL3 with the stiffest 
EOS generates the largest stellar radii. However, since the maximum neutron
star mass is sensitive to the EOS at higher densities, FSUGold 
(with $M_{\rm max}\!=\!1.72\,M_{\odot}$) becomes the softest beyond
$\mathcal{E}\!\sim\!200\,{\rm MeV/fm^{3}}$. 

Also displayed in Fig.\,\ref{Fig2} is the $P\!=\!\mathcal{E}$ line which 
marks the boundary beyond which an equation of state becomes 
\emph{ultrabaric}\,\cite{Glendenning:2000,Haensel:2007}. Being rooted 
on a Lorentz covariant framework, the predictions from all relativistic density 
functionals lie safely below the ultrabaric line. However, this is not the case 
for the R8--R11 profiles. Could this be an indication that such profiles violate 
\emph{causality}? That is, could the speed of sound exceed the speed 
of light in the high-density cores of these extremely compact stars? To 
test if a given EOS respects causality, we compute the speed of sound 
in the medium $c_{s}^{2}/c^{2}\!=\!dP/d{\mathcal E}$ and 
determine whether a point exists at which the EOS becomes superluminal. 
Such a point $(P_{c}, \mathcal{E}_{c})$ determines the central pressure 
and energy density of the heaviest neutron star that can be supported by 
the given EOS; beyond such a mass, the EOS required to support the 
star becomes acausal. 

\begin{figure}[ht]
\vspace{-0.05in}
\includegraphics[width=0.45\textwidth,angle=0]{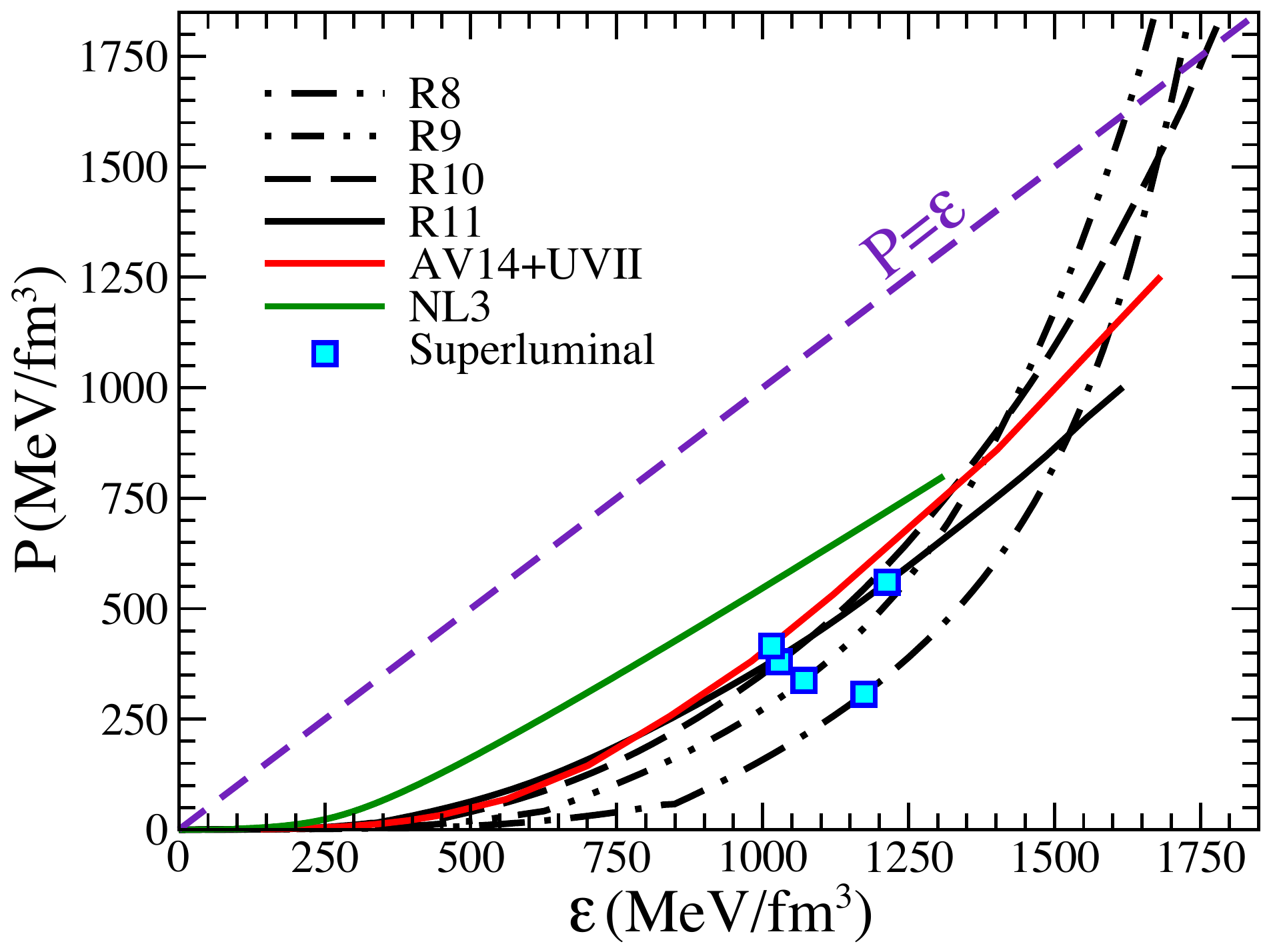}
\caption{(Color online) As in Fig.\,\ref{Fig2}, but now on a
reduced linear scale, we display some of the equations of 
state considered in this work; for simplicity, only NL3 is
included as a representative member of the relativistic
density functionals. With the exception of NL3, which is
causal at all densities, all other equations of state become
superluminal at the pressure and energy density indicated
by the blue squares. In turn, these values determine the
central pressure and energy density of the heaviest star 
that may be supported by a causal EOS.}
\label{Fig3}
\end{figure}

For clarity, we display in Fig.\,\ref{Fig3} the various equations of state 
on a reduced linear scale; only NL3 is included as a representative 
member of the relativistic density functionals. The figure clearly indicates 
that with the exception of NL3, all five equations of state become 
superluminal much before they cross into the ultrabaric region. The 
pressure and energy density at which the EOS becomes superluminal 
are depicted by the blue squares. In turn, these values determine the 
central pressure and energy density of the most massive neutron star 
that may be supported by a causal EOS; beyond this value the mass-radius 
relation becomes unphysical. We found that the value of this critical mass 
$M_{\rm crit}$ decreases rapidly as the common radius decreases. 
Specifically, although in the case of the R11 template we find 
$M_{\rm crit}\!=\!2.06\,M_{\odot}$, and thus consistent with the present 
$2M_{\odot}$ limit\,\cite{Demorest:2010bx,Antoniadis:2013pzd}, we obtain 
$M_{\rm crit}/M_{\odot}\!=\!1.83, 1.57, 1.26$, for R10, R9, and R8, respectively. 
Note that the microscopic model of Wiringa, Fiks, and Fabrocini violates
causality beyond $M_{\rm crit}\!\approx\!1.96\,M_{\odot}$\,\cite{Wiringa:1988tp}.
These values of $M_{\rm crit}$ define the lower boundary of the grey region 
labeled as ``Causality this work" in Fig.\,\ref{Fig1}. We thus conclude, within 
the scope of the adopted MR template, that in order to satisfy the current 
$2M_{\odot}$ limit, the stellar radius of a $1.4 M_{\odot}$ neutron 
star must exceed 10.7\,km. This represents the main finding of our work.

To place our newly developed constraint on the proper context, we 
display in Fig.\,\ref{Fig1} two other commonly used limits. The 
weakest of the two (labeled ``BH'') represents the black hole limit that 
precludes a neutron star of mass $M$ from having a radius $R$ equal 
to its Schwarzschild radius. This constraint has no impact 
on the mass-radius profiles considered in this work. A more stringent 
constraint (labeled ``L\&P'') follows from an analysis by Lattimer and
Prakash\,\cite{Lattimer:2006xb} that uses a realistic EOS up to a certain
value of the energy density that is then matched to a causally limited 
EOS. By doing so, they obtain a limit on the stellar compactness that 
may be written as $R\!\gtrsim\!2.83\,GM/c^{2}$. Although stronger than 
the BH constraint, this causality limit does not affect the recent analysis 
by Guillot and Rutledge\,\cite{Guillot:2014lla}. Indeed, the L\&P limit for 
a 9\,km neutron star is 2.16\,$M_{\odot}$. On the other hand, our analysis 
provides an even more stringent constraint of 1.57\,$M_{\odot}$, and 
pushes the minimum radius to 10.7\,km---barely consistent with 
the upper limit quoted in Ref.\,\cite{Guillot:2014lla}, but fitting comfortably 
within the $10.8^{+0.5}_{-0.4}$\,km range given in the very recent work 
by \"Ozel {\it et al.}\,\cite{Ozel:2015fia}.

\begin{figure}[ht]
\vspace{-0.05in}
\includegraphics[width=0.45\textwidth,angle=0]{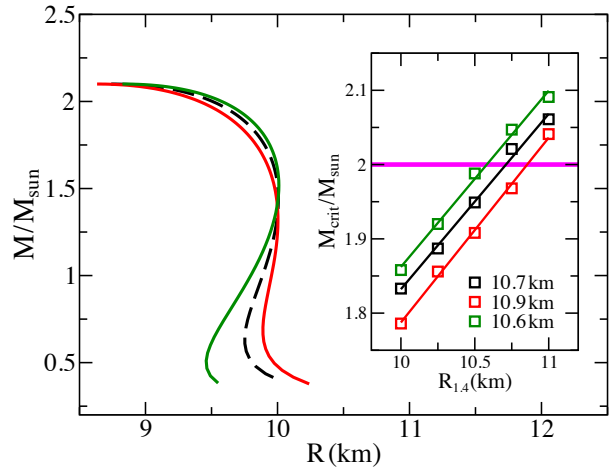}
\caption{(Color online) Sensitivity to small variations to the
standard MR template. Here the R10 template (dashed line)
is modified to produce two additional MR profiles, both with 
a radius of 10\,km for a $1.4M_{\odot}$ neutron star. The 
right portion of the plot displays the critical mass beyond 
which the underlying EOS becomes superluminal. Finally, 
the labels indicate the minimum radius required to support 
a $2M_{\odot}$ neutron star with a causal EOS of the given 
shape.}
\label{Fig4}
\end{figure}

Given that the MR profile is the only assumption made in this work, it 
is pertinent to ask how sensitive are our conclusions to the underlying 
shape. 
As shown in Fig.\,\ref{Fig4}, a constant-radius profile is characterized by a large 
(indeed infinite) derivative in the mass region of about 
$0.6\!\lesssim\!M/M_{\odot}\!\lesssim\!1.6$. If this shape is changed by
keeping the slope large but now negative (i.e., an increase in mass is 
accompanied by a reduction in the radius) then the compactness will 
increase and the underlying EOS will become superluminal even
earlier. However, if instead the slope becomes positive (i.e., an increase 
in mass is followed by an increase in radius) then it becomes harder to 
assess whether such an MR profile will generate an acausal EOS; such
a shape is displayed by the green solid line in Fig.\,\ref{Fig4}. Still, our 
results suggest that if variations to the standard template are not overly 
dramatic, our main conclusions remain valid. Indeed, the minimum 
stellar radius got shifted by only 0.1\,km; from 10.7 to 10.6\,km.

In summary, motivated by the latest progress in measuring neutron star radii, 
which combined with measurements of massive neutron stars provide critical
insights into the dense matter equation of state, we have examined whether
neutron stars may be as compact as recently suggested. To do so, we relied
on a powerful algorithm developed by Lindblom to obtain the equation of state 
from knowledge of the mass-radius relationship. That is, given a MR profile,
the underlying EOS may be obtained without any assumption on the nature or 
composition of the dense stellar core. Thus, all model dependence lies in the 
assumed MR profile, which in the present study was chosen to have a nearly 
constant radius shape. Using such a template, MR profiles were constructed
with a maximum mass of $2.1\,M_{\odot}$ and with common radii spanning the 
8 to 11\,km interval. Using Lindblom's algorithm, the equation of state 
associated with each of these MR profiles was obtained. Further, by imposing
causality, namely, by enforcing that the speed of sound be less than the speed 
of light, we obtained a stringent constraint on the maximum compactness of
neutron stars. Indeed, by demanding that a \emph{causal} EOS be able to 
support a 2$M_{\odot}$ neutron star, we obtained --- within the scope of the 
adopted template --- a lower limit on the 
stellar radius of a $1.4 M_{\odot}$ neutron star of 
$R_{\rm NS}^{\rm min}\!=\!10.7$\,km. Note that our result imposes
a \emph{lower} limit on stellar radii. In contrast, recent observational studies 
are placing \emph{upper} limits on neutron star radii. We trust that such a 
theoretical-observational synergy will continue to prove beneficial in the coming 
years in order to determine the dense matter equation of state.

This material is based upon work supported by the U.S. Department of 
Energy Office of Science, Office of Nuclear Physics under Award Number 
DE-FD05-92ER40750.

\bibliography{./Compactness}

\end{document}